\documentclass{article}

\usepackage{arxiv}

\usepackage[T1]{fontenc}
\usepackage{graphicx}
\usepackage{amsmath}
\usepackage{amsfonts}
\usepackage{amssymb}
\usepackage{subcaption}
\usepackage{tikz}
\usetikzlibrary{arrows.meta}
\usepackage{caption}
\usepackage[percent]{overpic}

\usepackage{hyperref}       % hyperlinks
\usepackage{url}            % simple URL typesetting
\usepackage{booktabs}       % professional-quality tables
\usepackage{amsfonts}       % blackboard math symbols
\usepackage{nicefrac}       % compact symbols for 1/2, etc.
\usepackage{microtype}      % microtypography
\usepackage{lipsum}		% Can be removed after putting your text content
\usepackage{graphicx}
\usepackage{natbib}
\usepackage{doi}

\renewcommand{\.}{\hspace*{0.07em}}

\usepackage[normalem]{ulem}

\title{The Spatial and Temporal Resolution of Motor Intention in Multi-Target Prediction}

\author{ \href{https://orcid.org/0009-0001-0420-6933}{\includegraphics[scale=0.06]{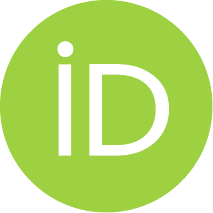}\hspace{1mm}Marie~D.~Schmidt}\thanks{
\url{https://lab.iossifidis.net/}        
        } \\
$^1$Robotics and BCI Laboratory, \\Institute of Computer Science,\\
Ruhr West University of Applied Science\\ 
Mülheim an der Ruhr, Germany \\
$^2$Institute for Neural Computation\\ Ruhr-University Bochum, Bochum, Germany\\
	\texttt{marie.schmidt@ruhr-uni-bochum.de} \\
	\And
	\href{https://orcid.org/0000-0002-9876-4396}{\includegraphics[scale=0.06]{orcid.pdf}\hspace{1mm}Ioannis Iossifidis} \\
$^1$Robotics and BCI Laboratory, \\Institute of Computer Science,\\
Ruhr West University of Applied Science\\ 
Mülheim an der Ruhr, Germany \\
	\texttt{iossifidis@hs-ruhrwest.de} \\
}

\begin{document}

% \author{Marie~D.~Schmidt\inst{1,2}\orcidID{0009-0001-0420-6933} \and
% Ioannis~Iossifidis\inst{1}\orcidID{0000-0002-9876-4396}}
% %\authorrunning{M.~D.~Schmidt et al.}

% % \institute{Institute of Computer Science, University of Applied Science Ruhr West, Mülheim an der Ruhr, Germany \and Institute for Neural Computation, Ruhr-University Bochum, Bochum, Germany
% % \email{marie.schmidt@ruhr-uni-bochum.de}\\
% % \url{https://lab.iossifidis.net/}
% }
%
\maketitle
\begin{abstract}
Reaching for grasping, and manipulating objects are essential motor functions in everyday life. Decoding human motor intentions is a central challenge for rehabilitation and assistive technologies. This study focuses on predicting intentions by inferring movement direction and target location from multichannel electromyography (EMG) signals, and investigating how spatially and temporally accurate such information can be detected relative to movement onset. We present a computational pipeline that combines data-driven temporal segmentation with classical and deep learning classifiers in order to analyse EMG data recorded during the planning, early execution, and target contact phases of a delayed reaching task.

Early intention prediction enables devices to anticipate user actions, improving responsiveness and supporting active motor recovery in adaptive rehabilitation systems. Random Forest achieves $80\%$ accuracy and Convolutional Neural Network $75\%$ accuracy across $25$ spatial targets, each separated by $14^\circ$ azimuth/altitude. Furthermore, a systematic evaluation of EMG channels, feature sets, and temporal windows demonstrates that motor intention can be efficiently decoded even with drastically reduced data. This work sheds light on the temporal and spatial evolution of motor intention, paving the way for anticipatory control in adaptive rehabilitation systems and driving advancements in computational approaches to motor neuroscience.

\end{abstract}

\section{Introduction}\label{Sec:Intro}
Reaching, grasping, and manipulating objects are fundamental components of human daily activity. These motor functions rely critically on the upper limb, highlighting the importance of preserving arm function for independence and interaction with the environment. While the neural decision-making and motor control are complex \cite{schmidt2026insights}, in this study, we focus specifically on intention prediction, the ability to infer the direction and target of a movement from peripheral signals, such as electromyography (EMG).

A key question in this context is the spatial resolution that can be achieved using EMG signals: how precisely can the intended movement be classified in space, and from what point in time relative to movement onset can the target be predicted? Previous studies suggest that neural signals recorded via EEG may reflect a general “go” signal, but provide limited information about the specific movement direction. Previous work has demonstrated that measurable EMG activity emerges as early as approximately $50$ ms prior to movement onset, indicating early activation of motor commands \cite{deecke1969distribution} (Figure \ref{fig:ExperimentalSetup}d)). It remains an open question whether such pre-movement EMG activity also encodes spatial aspects of the upcoming movement, such as reach direction or endpoint. Understanding these limits is essential not only for basic neuroscience, but also for practical applications in rehabilitation and human–machine interfaces (HMI), such as prosthetic control and assistive robotics.

Early prediction of movement intention provides a decisive advantage because it allows HMI devices to anticipate the user’s actions rather than simply reacting. In rehabilitation, this capability enables adaptive systems, such as exoskeletons, robotic arms, or functional electrical stimulation devices, to support patients precisely when they intend to move, encouraging active participation and promoting more effective motor recovery. Further improving fluidity, reducing delays, and creating a more intuitive control experience.

To demonstrate this, we use a delayed reaching task investigate upper-limb movement. In this task, the spatial goal of the reach is revealed to the participant in advance, while the initiation of the movement is withheld until a go cue that occurs after a randomly varying delay. During this delay period, the target location is known, and the reaching movement can therefore be fully planned, yet its execution is temporally constrained. This paradigm ensures that any EMG activity observed before movement onset can be attributed to motor planning and preparatory processes rather than to actual movement execution. This experimental design allows us to examine whether EMG signals recorded during the preparatory phase contain sufficient information not only to predict the imminent onset of movement, but also to infer the intended movement direction or target location before execution begins.

First of all, as a baseline, we assess the classification performance across $5\times5$ spatial targets distributed by $14^\circ$ azimuth/altitude to determine how accurately EMG can resolve different movement directions based on the EMG signal from movement onset and initial target contact. To achieve this, we employ a Random Forest (RF) and Convolutional Neural Network (CNN) classifiers, suited for these tasks \cite{mora2021multi}.
We then systematically evaluate channel selection and feature reduction, assessing how the number of EMG channels and temporal features affects predictive performance. Furthermore, we analyze predictive accuracy across temporal windows, spanning pre-motion, early motion, late motion, and holding to determine how intention-related information evolves over time.
%
%
%
%%%%%%%%%%%%%%%%%%%%%%%%%%%%%%%%%%%%%%%%%%%%%%%%
%\clearpage

\section{Method}
In this section, we describe the experimental setup for the reaching task and the preprocessing steps applied to the recorded data, as well as the selection of EMG features.

\subsection{Setup}\label{sec:Setup}
\begin{figure}[h!]
\centering
% ---- Column 1: a stacked on b ----
\begin{minipage}[t]{0.2\linewidth}
\vspace{0pt}
\centering
\begin{overpic}[width=0.9\linewidth]{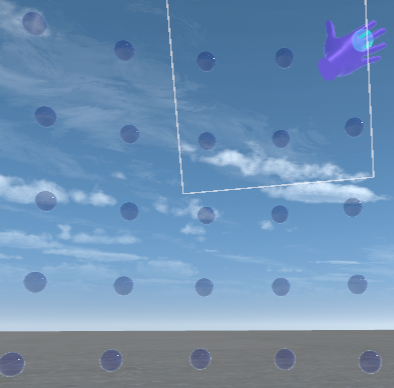}
    \put(-13,85){\normalsize a)}
\end{overpic}
\vspace{0.5em}
\begin{overpic}[width=0.9\linewidth]{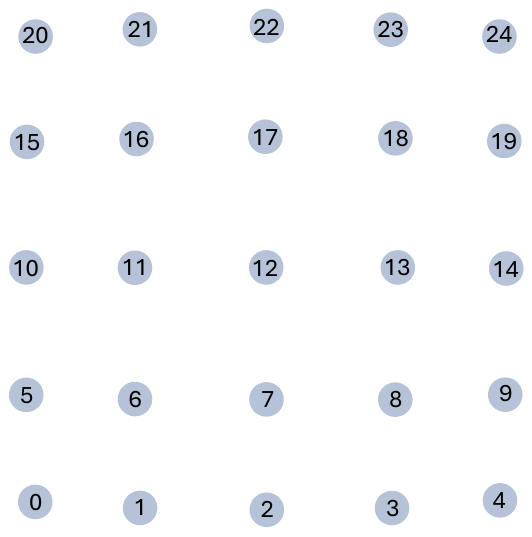}
    \put(-13,85){\normalsize b)}
\end{overpic}
\end{minipage}
\hfill
% ---- Column 2: c and d side by side ----
\begin{minipage}[t]{0.79\linewidth}
\centering
\begin{minipage}[t]{0.4\linewidth}
\vspace{0pt}
\centering
\begin{overpic}[width=0.9\linewidth]{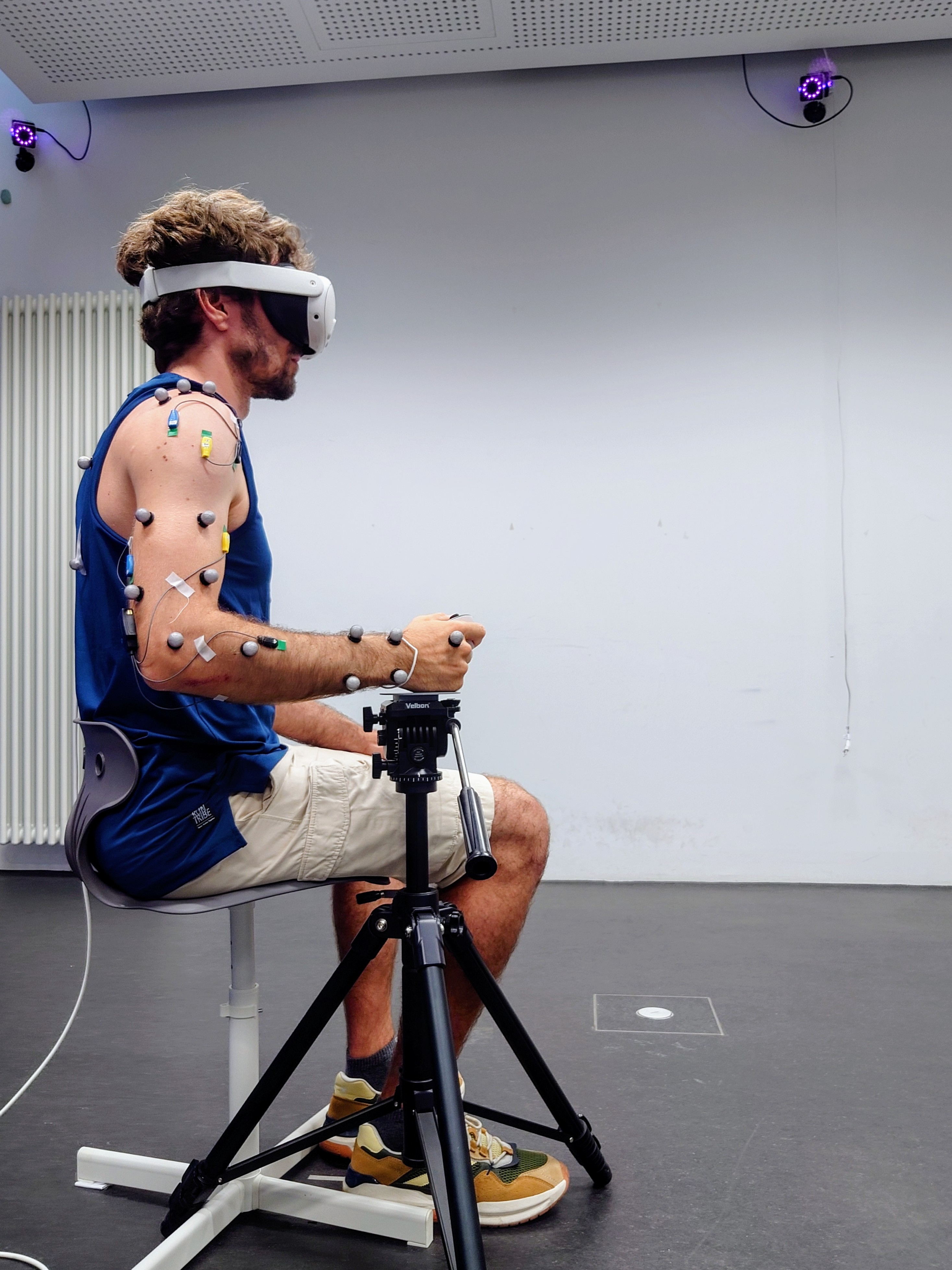}
    \put(-7,95){\normalsize c)}
\end{overpic}
\end{minipage}
\hfill
\begin{minipage}[t]{0.58\linewidth}
\vspace{0pt}
\centering
\begin{overpic}[width=1\linewidth]{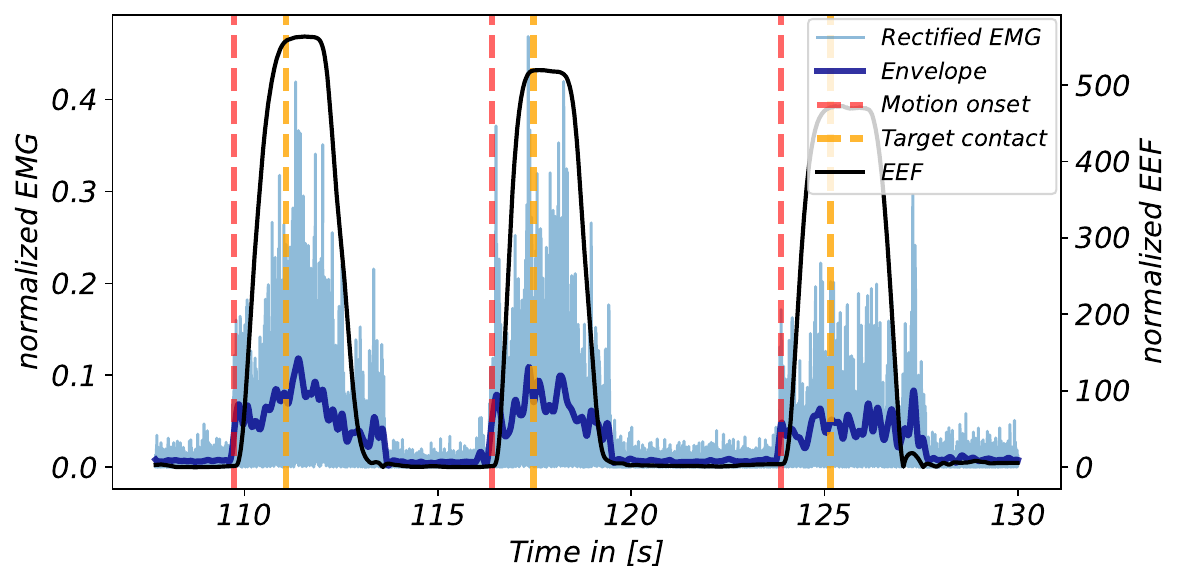}
    \put(-1,44){\normalsize d)}
\end{overpic}
$\phantom.$
\vspace{-0.8em}
\begin{overpic}[width=0.6\linewidth, trim = 0 0 0 0.4cm, clip]{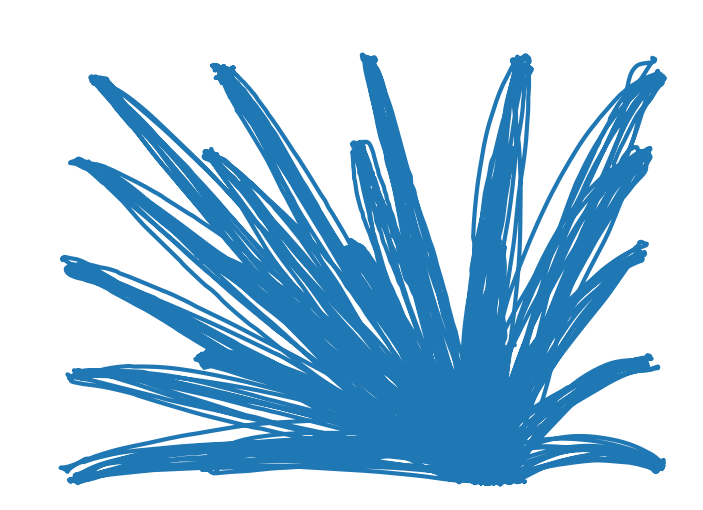}
    \put(-1,60){\normalsize e)}
\end{overpic}
\end{minipage}
\end{minipage}
% ---- Row 3: full width ----
\vspace{0em}
\begin{minipage}{\linewidth}
\vspace{0pt}
    \centering
    \begin{tikzpicture}[
        timeline/.style={thick, -{Stealth[length=3mm]}},
        state/.style={font=\small, align=center},
        sphere/.style={circle, minimum size=5mm, draw=none}]
    % Timeline
    \draw[timeline] (-1,0) -- (11,0);
    % State labels
    \node[state] (cue)     at (0,0.3)  {Cue};
    \node[state] (go)      at (2.5,0.3)  {Go};
    \node[state] (reach)   at (6.5,0.3)  {Reached};
    \node[state] (hold)    at (10.5,0.3) {Hold};
    % Vertical connectors
    \draw (0,0) -- (cue.south);
    \draw (2.5,0) -- (go.south);
    \draw (6.5,0) -- (reach.south);
    \draw (10.5,0) -- (hold.south);
    % Dashed intervals
    \draw[dashed] (0,0) -- (3,0);
    \draw[dashed] (2.5,0) -- (7,0);
    \draw[dashed] (6.5,0) -- (11,0);
    % State indicators
    \node[sphere, fill=orange!70, opacity=0.6] at (0,-0.3) {};
    \node[sphere, fill=green!70,  opacity=0.6] at (2.5,-0.3) {};
    \node[sphere, fill=blue!30,   opacity=0.7] at (6.5,-0.3) {};
    \node[sphere, fill=gray!70,   opacity=0.6] at (10.5,-0.3) {};
    \end{tikzpicture}
    \put(-350,22){f)}
\end{minipage}
\vspace{-0.2cm}
\caption{Experimental setup and task structure: a) VR environment from the participant’s perspective. b) target classes. c) placement of motion capture markers and EMG electrodes. d) EMG and End-EFfector trajectory, with vertical lines at task events. e) EEF trajectory. f) task event timeline with colored markers indicating task states.}
\label{fig:ExperimentalSetup}
\end{figure}
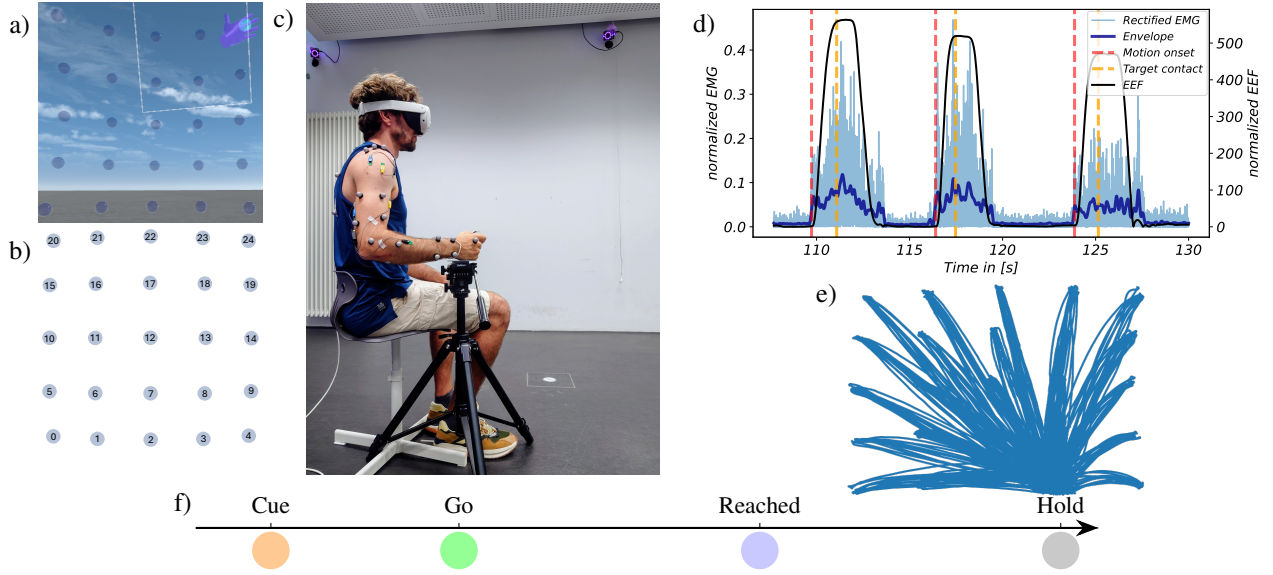

The delayed reaching task is implemented in Unity and presented through an Oculus Quest $2$ VR headset. Participants are instructed to reach toward the targets, represented as virtual spheres within the virtual environment. The task involves a $5 \times 5$ grid of spheres equally spaced with $14^\circ$ on a sphere surface, see Figure \ref{fig:ExperimentalSetup}a,b) with a radius adjusted to the subject’s arm length, which is roughly between $500$ mm and $650$ mm, see Figure \ref{fig:ExperimentalSetup}e).

This allows the spheres to be reached with an outstretched arm without using the upper body to rotate. During the task, the participant sits on a chair with a small backrest. To the right is a platform on which the arm can rest at a $90$-degree angle during the task, see Figure \ref{fig:ExperimentalSetup}c). The task logic works as follows: In the neutral layout, i.e.\ before start and between trials, all spheres are transparently present. To instruct the participant which sphere to reach for, a random sphere starts to illuminate in orange, and an additional stereo beep sound appears. The participant needs to fixate on the sphere with their sight. Once the target is seen, after a short randomized delay, it becomes active: the color of the sphere changes to green, and the participant is allowed to start the reach. This is therefore a delayed reaching task. Once the participant has reached the sphere, it changes color to blue and must be held for a further second, see Figure \ref{fig:ExperimentalSetup}f). After that, the sphere reverts to the neutral layout, and the participant needs to return to the platform before the next iteration can begin. During the task, the participant’s hand is visualised with a virtual purple hand, see Figure \ref{fig:ExperimentalSetup}a). The task comprises three sessions, each consisting of six repetitions, such that each sphere is reached six times in every session and $18$ times in total per participant.

The study involving human subjects was reviewed and approved by the Ethics Committee of the Ruhr-University Bochum. All methods were performed in accordance with the relevant guidelines and regulations. All participants gave their written consent.

The task is recorded simultaneously with EMG and positional data. The EMG is recorded via the Delsys Trigno system at $2000$ Hz with $10$ electrodes at the trapezius, latissimus, pectoralis, deltoid (anterior, posterior, and lateral), biceps and triceps, and the hand extensor and flexor, see Figure \ref{fig:ExperimentalSetup}c). Electrode placement and preparation are performed in accordance with the SENIAM standard \cite{konrad2005emg}. Prior to the task, the maximum voluntary contraction (MVC) for each muscle is performed. The EMG signal is filtered with a Butterworth $5-500$ Hz filter and with an adaptive filter at $50$ Hz. The signal is further rectified and normalized. The envelope is obtained by applying a zero-phase, two-sided low-pass filter with a $5$ Hz cutoff frequency. The positional data is recorded via Vicon Nexus with six Vicon cameras at $100$ Hz using a modified Southampton Upper Limb \cite{Warner2023ProCalc} with $20$ markers on the right arm and thorax, see Figure \ref{fig:ExperimentalSetup}d,e). The Vero cameras are set up in a semi-circle in front of the task area.
%
%
%
%%%%%%%%%%%%%%%%%%%%%%%%%%%%%%%%%%%%%%%%%%%%%%%%
%\clearpage

\subsection{EMG Feature Extraction}\label{sec:EMG Feature Extraction}
EMG features commonly used for upper-limb intention decoding are extracted from the time, frequency, and time–frequency domains \cite{phinyomark2018feature}, \cite{corvini2025emg} and let $x_1,\cdots,x_N$ denote the discrete EMG signal $x_n$ within a window of length $N$.

\textbf{Time-domain features:} For each channel, we compute the mean absolute value (MAV),
root mean square (RMS),
waveform length (WL),
variance (VAR),
integrated EMG (IEMG),
and Slope (SL)
as well as the minimum and maximum values
\begin{align*}
\mathrm{MAV} &= \frac{1}{N}\sum_{n=1}^N|x_n|\,, &
\mathrm{RMS} &= \sqrt{\frac{1}{N}\sum_{n=1}^Nx_n^2}\,, &
\mathrm{WL}  &= \sum_{n=1}^{N-1}|x_{n+1}-x_n|\,, \\
\mathrm{VAR} &= \frac{1}{N-1}\sum_{n=1}^N(x_n-\overline{x})^2,&
\mathrm{IEMG} &= \sum_{n=1}^N|x_n|\,, &
\mathrm{SL} &= x_N - x_1\,.
%\mathrm{MIN} &= \min_n x_n\,, &
%\mathrm{MAX} &= \max_n x_n\,.
\end{align*}

\textbf{Frequency-domain features} are derived from the power spectral density $P_j$ estimated using Welch’s method, where $P_j$ and $f_j$ are the amplitude spectrum and the frequency of spectrum at frequency bin $j$.
We extracted the mean frequency (MNF), median frequency (MDF), peak frequency (PKF), spectral entropy (SE), total power (TP), and relative band power (BP$_b$):
\begin{align*}
\mathrm{MNF} &= \frac{\sum_{j=1}^M f_j\. P_j}{\sum_{j=1}^M P_j}\,,&
\mathrm{MDF}&:\;\sum_{f\leq\mathrm{MDF}} P_j=\frac{1}{2}\sum_{j=1}^M P_j\,,&
\mathrm{PKF} &= \arg\max_f P_j\,, \\
\mathrm{SE} &= -\sum_{j=1}^M  \widetilde{P}_j\log \widetilde{P}_j\,,&
\mathrm{TP} &= \sum_{j=1}^M P_j\,,&
\mathrm{BP}_b &= \frac{\sum_{f\in b}P_j}{\sum_{j=1}^M P_j}\,.
\end{align*}
where $ \widetilde{P}j$ denotes the normalized PSD and we have $b_\alpha=[8,12]$ $b_\beta=[15,35]$, $b_\gamma=[35,60]$, $b_\delta=[60,500]$~Hz as the frequency bands of interest.

\textbf{Wavelet-domain features} are obtained using a discrete wavelet transform (db4, $L=4$).
For wavelet coefficients $c_{l,n}$ at level $l=1,\cdots,4$, we compute the wavelet energy E$_l$, relative energy $\widetilde{\rm E}_l$, and wavelet entropy WE
\begin{equation*}
    \rm{E}_l=\sum_{n=1}^N c_{l,n}^2\,,\qquad 
    \widetilde{\rm E}_l=\frac{E_l}{\sum_{k=1}^L E_k}\,,\qquad
    \mathrm{WE} = -\sum_{l=1}^{L}  \widetilde{E}_l \log  \widetilde{E}_l\,.
\end{equation*}
All features are computed independently for each EMG channel and concatenated into a single feature vector.

\subsection{Random Forest Classifier}
A Random Forest is an ensemble learning method that combines the predictions of many decision trees, where each tree is trained on a random subset of the data and a random subset of features, and the final prediction is obtained by majority voting for classification \cite{Breiman2001RF}. Random Forest classifiers are often selected as a baseline model due to their strong performance on high-dimensional, heterogeneous feature spaces and their robustness to noise and redundant information \cite{fernandez2014we}. EMG-derived feature representations typically contain correlated and partially redundant features arising from muscle synergies and co-contraction patterns. Random Forests naturally handle such redundancy through feature subsampling and ensemble averaging, reducing overfitting while maintaining discriminative power.

\subsection{CNN Classifier} \label{sec:CNN}
Convolutional Neural Networks (CNNs) have shown strong performance in EMG-based movement classification by automatically learning hierarchical spatio-temporal feature representations from raw or minimally processed signals \cite{mora2021multi}. In particular, one-dimensional CNNs effectively capture local temporal dependencies in EMG signals while maintaining computational efficiency. Therefore, we adopted a similar 1D CNN architecture that consists of three consecutive 1D convolutional layers, with $32$ filters and a kernel size of $3$. Each convolutional layer is followed by batch normalization and a ReLU activation function. Temporal dimensionality is reduced by max pooling, followed by global average pooling. Two fully connected layers (128 units each) with dropout of $0.2$ are used.

\subsection{Statistics}
We used a Wilcoxon rank-sum test, which is a non-parametric test used to determine if there is a significant difference between the medians of two independent, non-normally distributed samples. To account for multiple comparisons, p-values were adjusted using the Bonferroni correction.
%
%
%
%%%%%%%%%%%%%%%%%%%%%%%%%%%%%%%%%%%%%%%%%%%%%%%%
%\clearpage
\section{Results}
We present the outcomes of the muscle contribution analysis, channel and feature reduction, and their effects on classification performance, together with an analysis of time-segment contributions. In addition, we examine the relevance of pre-motion EMG activity for early and accurate motion discrimination.

\subsection{Spatial resolution Baseline Classification}
\begin{figure}[h!]
% \centering
\vspace{0em}
\begin{overpic}[width=.9\linewidth]{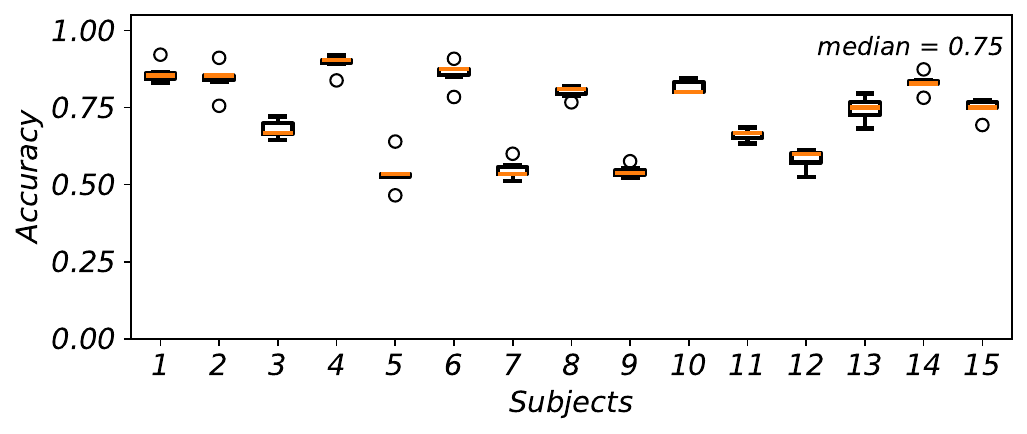}
    \put(-1,35){\normalsize a)}
\end{overpic}
\label{fig:RF_results_f28_c10_w1_nice}
\begin{minipage}[t]{0.45\linewidth}
\vspace{0pt}
\centering
% \vspace{0em}
% \begin{overpic}[width=\linewidth]{RF_results_f28_c10_w1_nice.pdf}
%     \put(-1,35){\normalsize a)}
% \end{overpic}
% \label{fig:RF_results_f28_c10_w1_nice}

\vspace{0pt}
\centering
\begin{overpic}[width=\linewidth, trim = 0 0 0 0.7cm, clip]{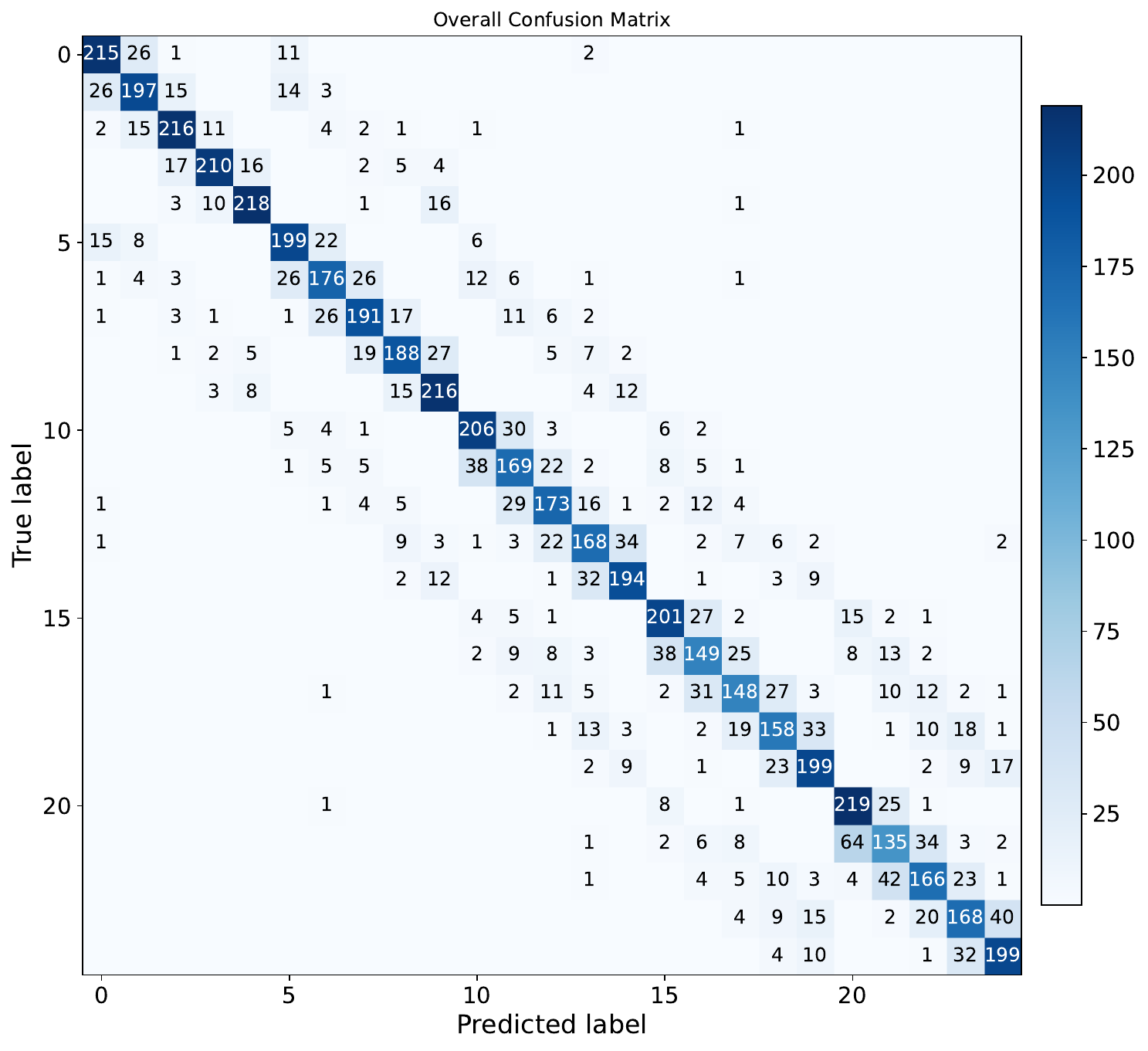}
    \put(0,83){\normalsize b)}
\end{overpic}
\label{fig:Confusion_Matrix_RF_Subject44_Datafeatures_onefull_Channels10_overall}
\end{minipage}
\hfill
\begin{minipage}[t]{0.45\linewidth}

\label{tab:optuna}
\renewcommand{\arraystretch}{2}
\centering\captionof{table}{Random Forest optimization.}
\begin{tabular}{ll}
\hline
\textbf{Hyperparameter} & \textbf{Value} \\
\hline
Number of estimators  & 500 \\
Maximum tree depth  & 10 \\
Minimum samples for split  & 2 \\
Minimum samples per leaf  & 3 \\
Maximum features  & ‘sqrt’ \\
\hline
\end{tabular}
\label{tab:optuna}
\renewcommand{\arraystretch}{1}
% \vspace{0em}
% \begin{overpic}[width=\linewidth]{RF_results_f28_c10_w1_nice.pdf}
%     \put(-1,35){\normalsize a)}
% \end{overpic}
% \label{fig:RF_results_f28_c10_w1_nice}

\end{minipage}
\vspace{-1em}
\caption{Random Forest: a) prediction for all participants based on the entire reach and all channels and features. b) confusion matrix across all subjects.}
\label{fig:baselineclassi}
\end{figure}

The Random Forest classifier is optimized using Optuna with a Tree-structured Parzen Estimator (TPE) sampler Table \ref{tab:optuna}. Model performance is evaluated using $5$-fold cross-validation with an $80/20$ train–test split in each fold. Separate subject-specific models are trained to account for inter-subject differences. As an initial benchmark, a baseline configuration is evaluated to provide an overall indication of classification performance. This baseline utilized all available EMG channels (Section \ref{sec:Setup}), a single temporal window spanning the entire reaching movement, and the complete set of all $28$ extracted features (Section \ref{sec:EMG Feature Extraction}). Under this configuration, the median classification accuracy across subjects is $75\%$, which means we can resolve up to $14$ degree. Notably, the results exhibited pronounced inter-subject variability, with individual accuracies ranging from approximately $90\%$ to $50\%$ for subjects, see Figure~\ref{fig:baselineclassi}a). Interestingly, subjects with lower accuracy between $50\%$ and $60\%$, i.e.\ number $5$, $7$, $9$, $11$, and $12$, show these similarities also for other Classification approaches.

\enlargethispage{1.3cm}

The confusion matrix illustrates the relationship between the predicted and true class labels. As expected, given the overall high classification accuracy, the diagonal elements dominate, indicating a high rate of correct predictions (Figure \ref{fig:baselineclassi}b)). Misclassifications predominantly occur between spatially adjacent targets, typically involving neighboring classes to the left, right, above, or below the true target location. This pattern suggests that classification errors are mainly confined to nearby spatial classes rather than distant targets, reflecting a graded spatial encoding of movement direction in the EMG signals (Figure~\ref{fig:baselineclassi}b)).
We subsequently aim to systematically reduce the data dimensionality to identify the EMG channels and feature subsets that contribute most to classification performance.

\subsubsection{Resolution Classification with fewer targets}
\begin{figure}[h!]
\centering
\begin{minipage}{0.25\linewidth}
\vspace{0pt}
\centering
\begin{overpic}[width=\linewidth]{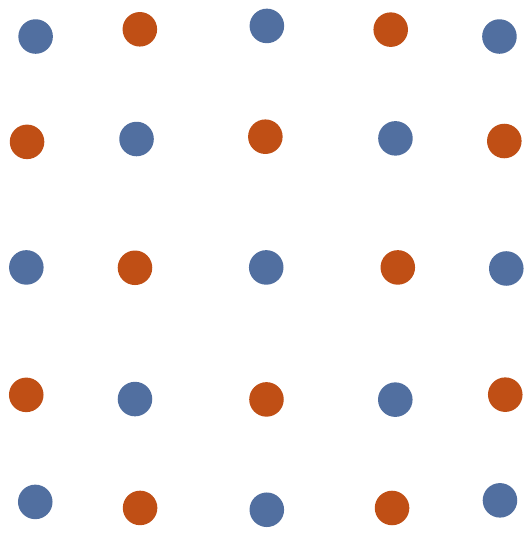}
    \put(-8,90){\normalsize a)}
\end{overpic}
\label{fig:kugelgrid_dark12}
\end{minipage}
\hfill
\begin{minipage}{0.73\linewidth}
\vspace{0pt}
\centering
\begin{overpic}[width=\linewidth]{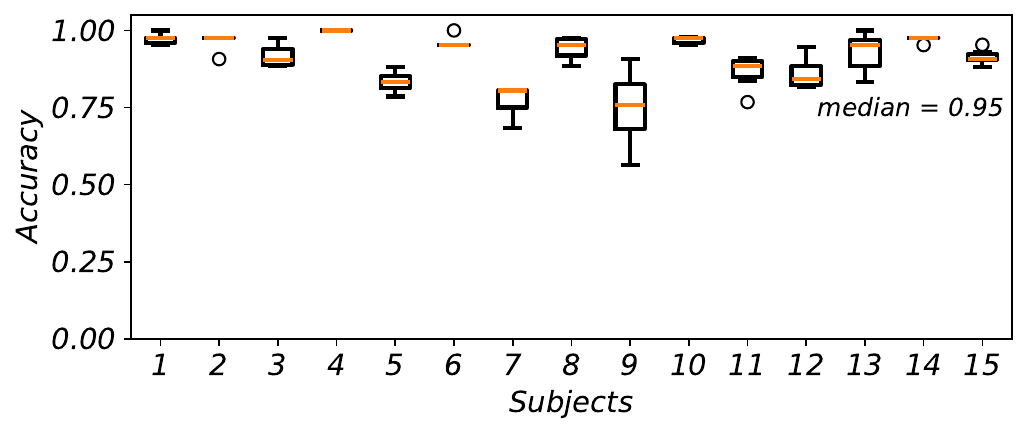}
\label{fig:RF_results_f8_c7_w8_nice}
    \put(0,37){\normalsize b)}
\end{overpic}
\end{minipage}
\caption{Random forest for $12$ targets: a) target distribution c) performance per subject.}
\label{fig:rf_12class}
\end{figure}

The general task includes $5\times5$ targets, each at a distance of $14^\circ$ azimuth/altitude. We can classify this with an accuracy around $75\%$. Reducing the number of targets, and therefore the number of classes, should improve the prediction and enable us to narrow down the spatial resolution (Figure~\ref{fig:rf_12class}a)). The scenarios where every second target is omitted, leaves us with $12$ or $13$ targets (Figure~\ref{fig:baselineclassi}b)). In these cases, we achieve a median score across participants of $95\%$ (Figure~\ref{fig:rf_12class}c)). This leads to a significant improvement in classification performance, demonstrating that $28^\circ$ azimuth/altitude can be predicted very well. However, we still observe decreased performance for the same weak subjects. All the other subjects are now close to $100\%$ performance. Again, misclassifications are mostly neighbouring classes.

\subsection{EMG channel importance}
\begin{figure}[h!]
\centering
\begin{minipage}{0.5\linewidth}
\centering
\captionof{table}{EMG channel relevance for single channel and leave one channel out.}
\begin{tabular}{ll|ll}
\hline
\textbf{Channel} & \textbf{Muscle} & \textbf{Single} & \textbf{Leave}\\
\hline
0 & triceps brachii      & 0.21 & 0.77\\
1 & biceps brachii       & 0.19 & 0.76\\
2 & wrist flexor         & \textbf{0.10} & 0.76\\
3 & wrist extensor       & \textbf{0.12} & 0.77\\
4 & deltoid posterior    & 0.25 & 0.75\\
5 & deltoid lateral      & 0.37 & 0.75\\
6 & deltoid anterior     & 0.28 & 0.77\\
7 & pectoralis           & 0.31 & \textbf{0.61}\\
8 & trapezius            & \textbf{0.15} & 0.76\\
9 & latissimus dorsi     & 0.26 & 0.76\\
\hline
\end{tabular}
\label{tab:channel_count}
\end{minipage}
\hfill
\begin{minipage}{0.49\linewidth}
\centering
\includegraphics[width=\linewidth]{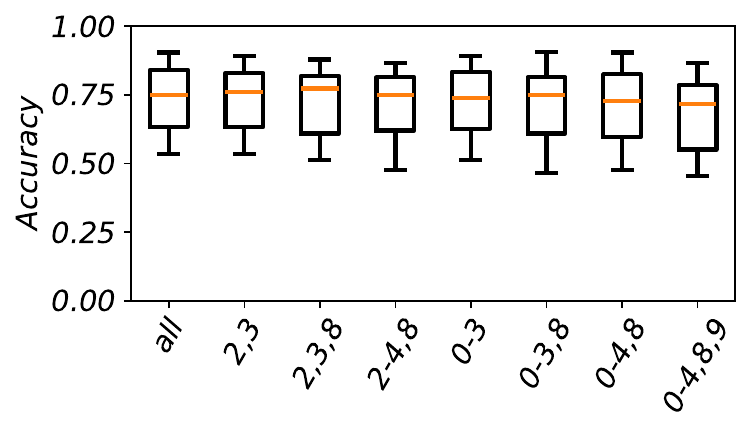}
\caption{Random Forest performance with different numbers of leave-out channels.}
\label{fig:RF_channel_results}
\end{minipage}
\end{figure}

We recorded electromyographic (EMG) signals from ten muscles that are primary contributors to the executed reaching motions, see Section \ref{sec:Setup}. However, some of these muscles may provide redundant information due to co-contraction or antagonistic activation patterns. Therefore, we analyze the relative contribution of each muscle to the prediction performance to identify the most informative channels (Table \ref{tab:channel_count}) and to assess how channel selection affects model accuracy.

%\clearpage
To quantify muscle relevance, we measure its performance alone as a single input, as well as the effect of excluding it as a leave-out channel (Table \ref{tab:channel_count}).
This analysis reveals that three channels: the wrist flexor and extensor as well as the trapezius, have the lowest contribution to the prediction task. Consistent with this finding, the RF model achieved a comparable median accuracy of approximately $75\%$ when excluding these and only using seven channels (Figures~\ref{fig:RF_channel_results}b)). In contrast, further reduction of the number of channels results in a decline in classification accuracy, indicating that while limited redundancy exists, most recorded muscles provide relevant information for the prediction task.

\subsection{Feature importance}
\begin{figure}[h!]
\centering
\begin{subfigure}[t]{0.55\linewidth}
\vspace{0pt}
\centering
\begin{overpic}[width=\linewidth]{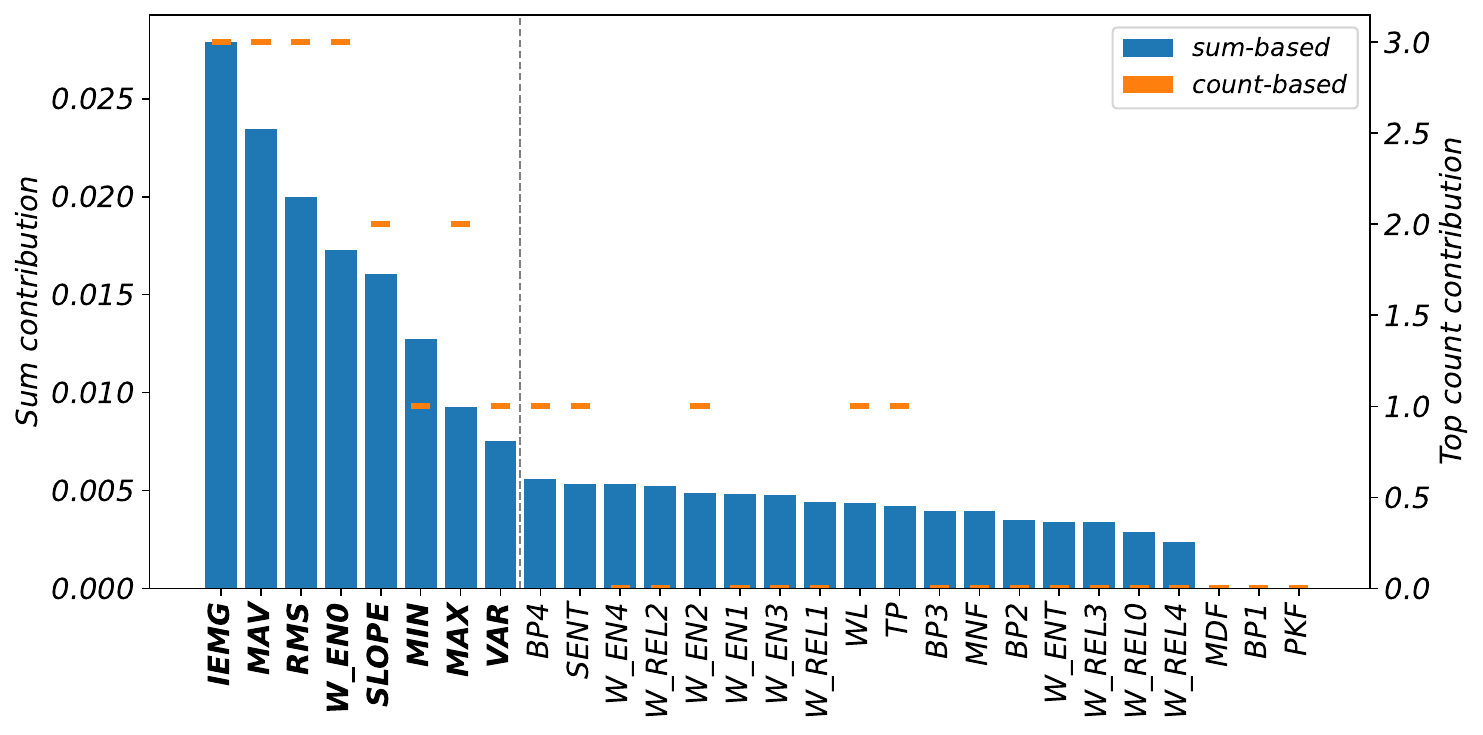}
    \put(-1,44){\normalsize a)}
\end{overpic}
\label{fig:RF_feature_count_nice}
\vspace{0em}
\begin{overpic}[width=0.7\linewidth]{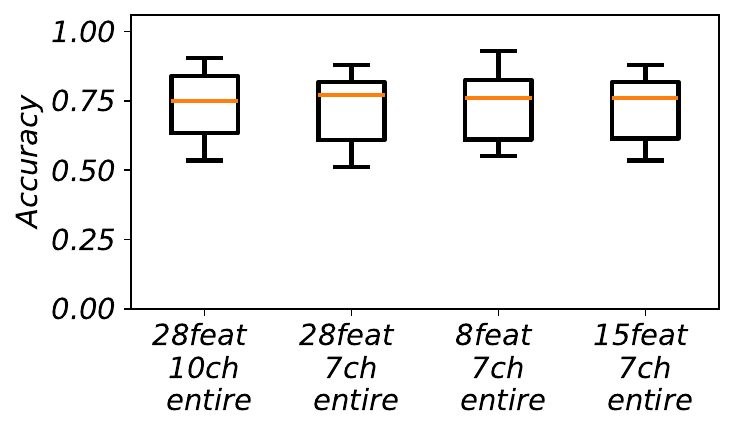}
    \put(0,50){\normalsize b)}
\end{overpic}
\label{fig:RF_results_feature_channel_entire}
\end{subfigure}
\hfill
\begin{subfigure}[t]{0.44\linewidth}
\vspace{0pt}
\centering
\begin{overpic}[width=\linewidth]{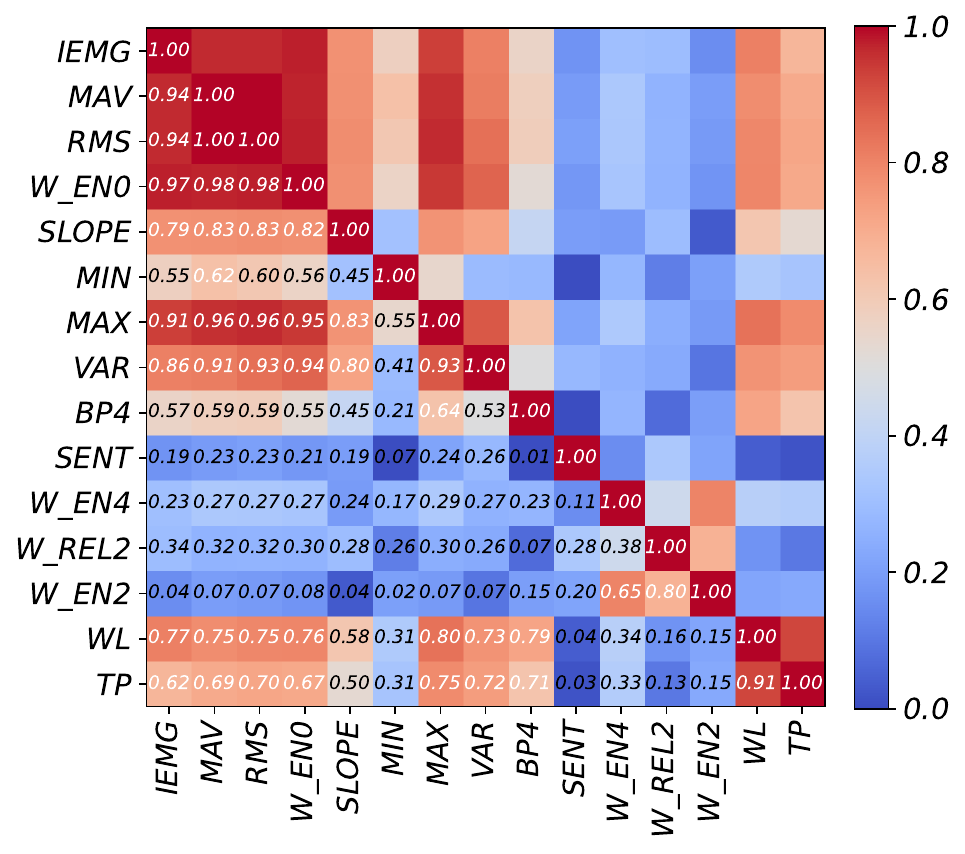}
    \put(0,80){\normalsize c)}
\end{overpic}
\label{fig:RF_feature15_ch7_Average_correlationmatrix_new}
\end{subfigure}
\caption{Random Forest feature analysis a) contribution of features. b) performance comparison between models based on different numbers of features and channels c) correlation matrix exemplary for $15$ features and $10$ channels.}
\label{fig:feature_importance}
\end{figure}

We extract $28$ different features from the time, frequency, and time–frequency domains, see Section \ref{sec:EMG Feature Extraction}. Examination of the feature correlation matrix (see Figure \ref{fig:feature_importance}c)) reveal that several features, particularly within the time-domain group, show high inter-feature correlations. However, RF classifiers are generally robust to such redundancy, as correlated features do not substantially degrade performance as long as relative importance is high. Thus, to further refine the feature set, we analyze the relative importance of each feature in terms of its contribution to prediction accuracy (Figure \ref{fig:feature_importance}a)). Based on this analysis, we select $8$ of the original $28$ features (Figure \ref{fig:feature_importance}), retaining all time-domain features except waveform length (WL) and including wavelet entropy, while excluding all frequency-domain features and most time–frequency features. Eight features were chosen because this represents the smallest subset that preserves maximal classification accuracy; comparable performance was also observed with $15$ features. Furthermore, although the correlation matrix indicated the presence of less strongly correlated features, their inclusion did not yield measurable improvements in predictive performance. After this feature selection and reduced channel amount, the median performance remains at $76\%$ while the overall amount of features is drastically reduced.

\subsection{Temporal resolution}
\begin{figure}[h!]
\centering
\begin{minipage}[t]{0.28\linewidth}
\vspace{0pt}
\centering
\captionof{table}{Window wise permutation importance.}
\begin{tabular}{ll}
\hline
\textbf{Window} & \textbf{Value} \\
\hline
1 & 0.006 \\
2 & 0.025 \\
3 & 0.031 \\
4 & 0.03 \\
5 & 0.024 \\
6 & 0.032 \\
7 & \textbf{0.059} \\
8 & \textbf{0.1} \\
\hline
\end{tabular}
\label{tab:window_count}
\end{minipage}
\hfill
\begin{minipage}[t]{0.7\linewidth}
\vspace{0pt}
\centering
\includegraphics[width=\linewidth]{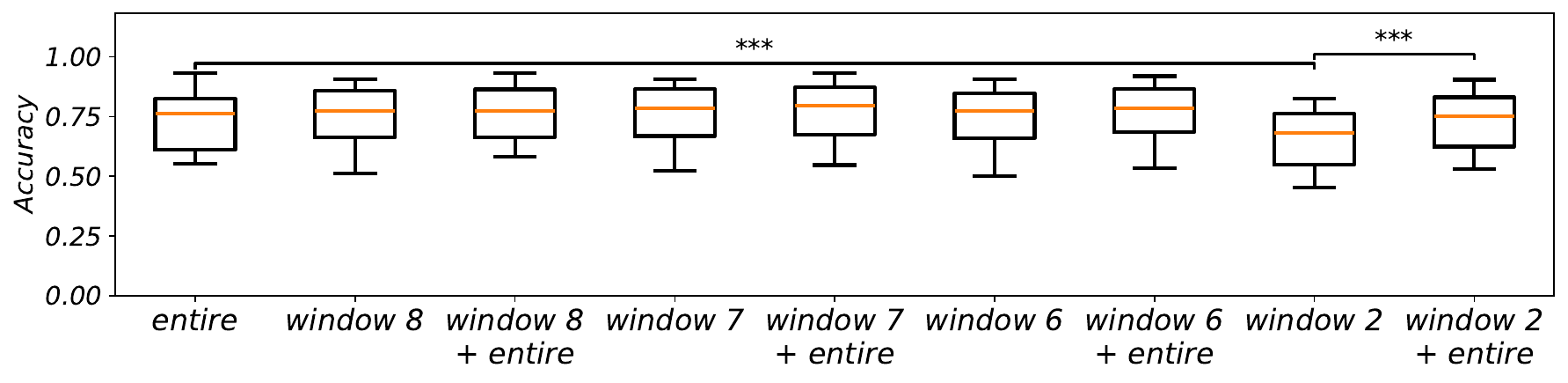}
\captionof{figure}{Performance for different temporal windows configurations. \textit{Entire} denotes a single window covering the reach, while the numbers indicate the number of temporal windows used. (Wilcoxon test: entire vs w2: $W = 1.000$, $p = 0.0001$, w2 vs w2+entire: $W = 0.000$, $p = 0.0001$, with Bonferroni correction).}
\label{fig:RF_results_feat8_ch7_windows}
\end{minipage}
\end{figure}

Up to now, all features have been calculated over the whole movement. We now want to address the question which temporal segment of the reach contributes most to classification performance? To address this question, the reaching movement is segmented from motion onset to initial target contact into eight non-overlapping temporal windows, each with a duration of approximately $200$ ms, which is recommended for classification \cite{irastorza2017design}. For each temporal windows we compute $8$ feature for $7$ channel each as discussed above. This windowing strategy provides a fine-grained temporal representation of the muscle activation patterns throughout the reach while preserving their chronological structure. Using this representation, we compute window-wise permutation importance to quantify the contribution of each temporal segment to classification performance. The results, summarized in Table \ref{tab:window_count}, indicate a clear temporal structure of the decisive factors. Specifically, the late time windows exhibit the highest contribution, with windows $7$ and $8$ showing substantially larger importance values than earlier windows. These windows correspond to approximately the final $400$ ms before target contact, suggesting that muscle activation patterns immediately preceding contact contain the most class-discriminative information.

Building on this observation, we evaluate the classification performance as a function of the selected temporal windows. Figure \ref{fig:RF_results_feat8_ch7_windows} reports accuracy for different window configurations. The highest performance is achieved when seven time windows are included leaving out the first temporal window, and further improves when the original window spanning the entire reach is added, yielding an accuracy of $80\%$. Accuracy decreases again with a further reduction in the number of temporal windows. This indicates that while late windows dominate the decision process, earlier temporal information still provides complementary cues that improve overall performance.

Overall, these results demonstrate that discriminative information is temporally concentrated toward the end of the reach, but that aggregating information across multiple time scales, including an entire window, yields the most robust classification.

\subsection{From what point in time can we predict targets?}
\begin{figure}[h!]
\centering
\begin{subfigure}[t]{\linewidth}
\vspace{0pt}
    \centering
    \begin{overpic}[width=\linewidth]{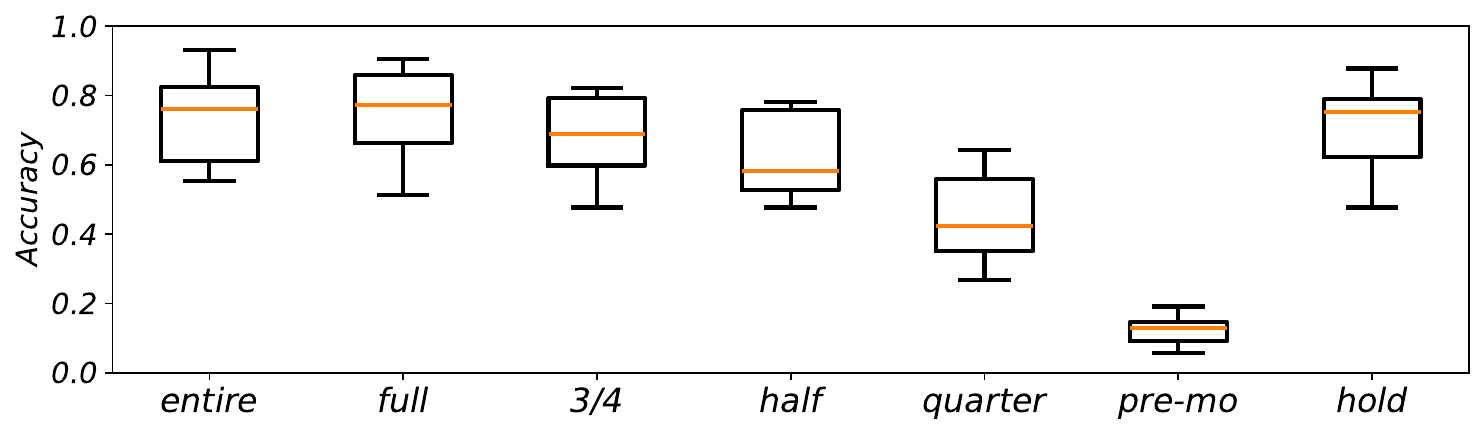}
        \put(-2.5,25){\normalsize a)}
    \end{overpic}
    \vspace{.0cm}
\end{subfigure}
\vspace{0.0cm}
\begin{subfigure}[t]{0.6\linewidth}
\vspace{0pt}
    \centering
    \begin{overpic}[width=\linewidth]{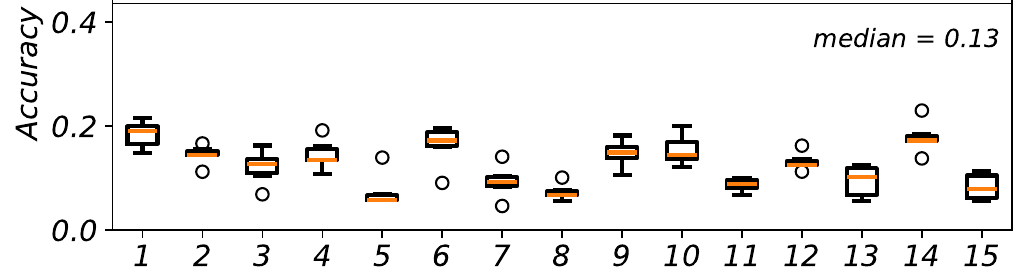}
        \put(-2.5,28){\normalsize b)}
    \end{overpic}
    \vspace{0.5cm}
    \begin{overpic}[width=\linewidth]{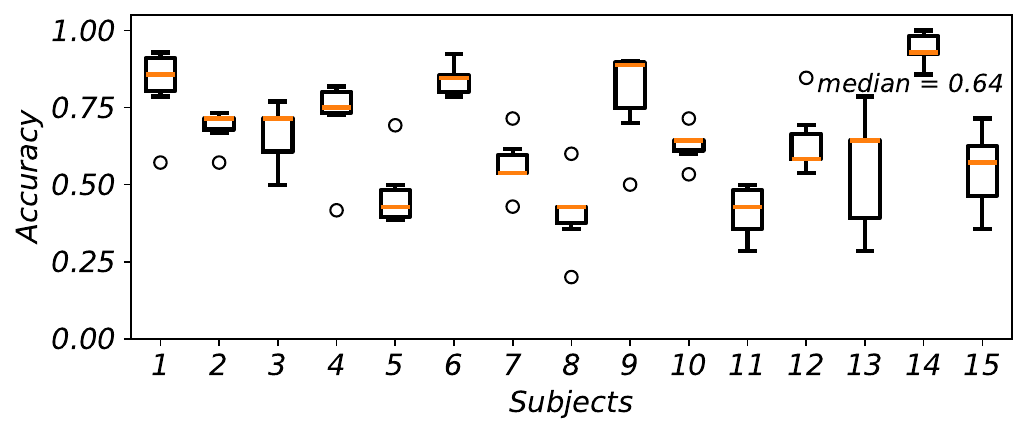}
        \put(-2.5,36){\normalsize c)}
    \end{overpic}
\end{subfigure}
\hfill
\begin{subfigure}[t]{0.39\linewidth}
\vspace{0pt}
    \centering
    \begin{overpic}[width=0.9\linewidth]{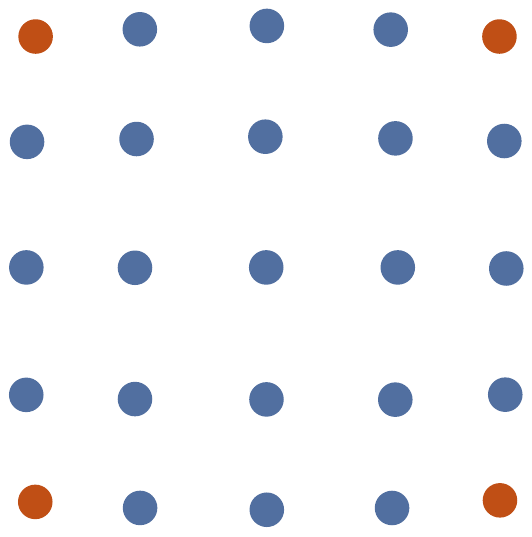}
        \put(-5,92){\normalsize d)}
    \end{overpic}
\end{subfigure}
\caption{Random Forest classification results: a) based on different points in time. b) pre-motion across participants. c) classification for a reduced number of classes $4$. d) class distribution.}
\label{fig:RF_results_sidebyside}
\end{figure}

We have demonstrated that multiple movement classes can be predicted reliably from EMG signals. Beyond identifying configurations that maximize classification performance, it is also clinically relevant to determine how early in time the intended movement direction can be inferred and what trade-offs in accuracy arise when predictions are made from partial temporal information. To this end, we systematically evaluate classification performance using progressively longer temporal segments, including the first quarter, first half, first three quarters, and full reach duration, as well as the post-contact hold phase. Additionally, we analyze pre-motion conditions in which the target is known but actual movement is constrained, see Figure \ref{fig:RF_results_sidebyside}a).

Prediction performance decreases systematically when the analysis is restricted to earlier temporal segments. Using only the first quarter of the reach results in an accuracy of $42\%$, reflecting lower discriminative information during early movement execution. Notably, the median classification accuracy during the pre-motion interval is $13\%$ across participants, despite the absence of observable movement, which is still three times better than a random selection. Figure \ref{fig:RF_results_sidebyside}b) also shows the prediction for individual participants displaying a wide range of predictive performance, ranging from $25\%$ to $10\%$, depending on the participant.

We further aim to determine the resolution limit of our approach, specifically the minimum number of classes to which the problem must be reduced in order to achieve reliable classification performance. By reducing the number of classes to $4$, i.e.\ only considering the corner targets, (Figure \ref{fig:RF_results_sidebyside}b)) the median classification accuracy across subjects increases to $64\%$ Figure \ref{fig:RF_results_sidebyside}c). However, substantial inter-subject variability persisted, with some participants achieving considerably higher performance than others. This finding suggests that preparatory muscle activation already encodes information about the intended target before movement onset, which has important implications for rehabilitation, as it enables anticipatory, intention-driven assistance rather than purely reactive control, thereby supporting more natural and effective motor recovery and daily-life interaction.

\subsection{CNN Classification}
\begin{figure}[h!]
    \centering
    \begin{overpic}[width=0.56\linewidth]{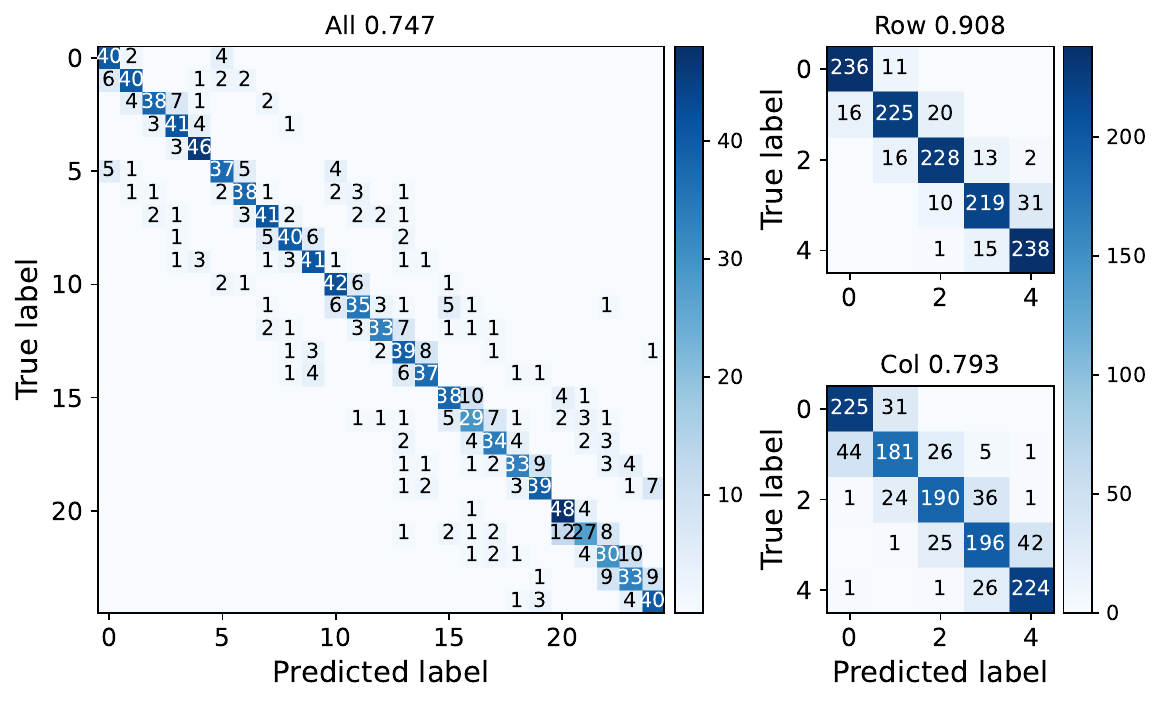}
    \put(0,52){\normalsize a)}
    \end{overpic}
    \hfill
    \begin{overpic}[width=0.201\linewidth]{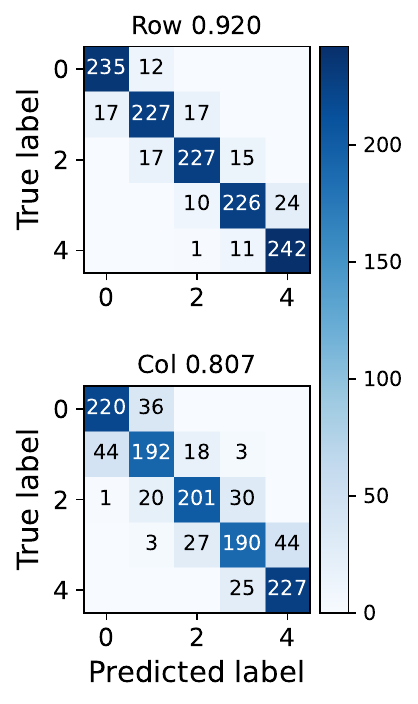}
    \put(-5,85){\normalsize b)}
    \end{overpic}
    \hfill
    \begin{overpic}[width=0.201\linewidth]{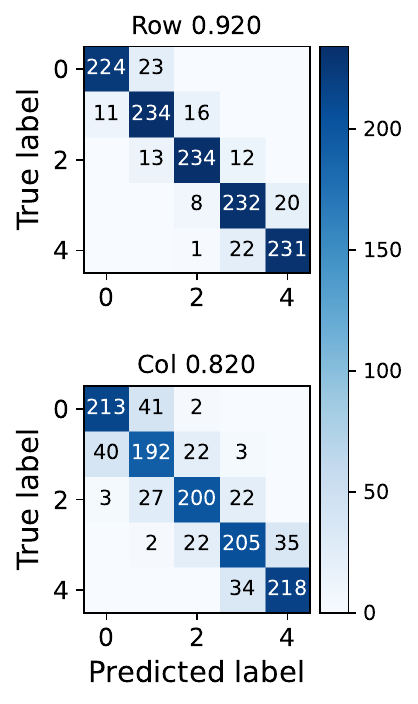}
    \put(-5,85){\normalsize c)}
    \end{overpic}
    \caption{Comparison of CNN confusion matrices: a) predicting 25 targets. b) two outputs for column and row. c) two separate CNNs predicting either row or column.}
    \label{fig:all_CNN_confusion_matrices}
\end{figure}

Although the Random Forest classifier yields strong performance, it depends on predefined feature representations and does not explicitly model temporal dependencies within the EMG signal. Given the structured, time-varying nature of EMG and the coordinated activation across muscle channels, we next employ a CNN to learn spatiotemporal representations directly from the preprocessed EMG signal. Temporal convolutions enable the extraction of local activation patterns and inter-channel dependencies in a data-driven manner, reducing reliance on manual feature engineering. CNNs are inherently capable of handling multivariate outputs and can be generalized from simple classification to continuous regression models. We examine three slightly different modeling strategies:
\begin{itemize}
    \item[$\bullet$] First, we implement a CNN that directly predicts all $25$ targets as a single output (Figure \ref{fig:all_CNN_confusion_matrices}a)) similar to the RF approach serving as the benchmark. 
    \item[$\bullet$] Second, we exploit the geometric structure of the task and decompose the classification problem into separate row and column components, enabling the CNN to predict both simultaneously within a single model using separate outputs (Figure \ref{fig:all_CNN_confusion_matrices}a)). 
    \item[$\bullet$] Third, we train two independent CNNs, one dedicated to row classification and the other to column classification (Figure \ref{fig:all_CNN_confusion_matrices}b)), maximizing flexibility by letting each network specialize on its respective axis.
\end{itemize}

Strategies 2 and 3 allow the networks to explicitly model the spatial relationships between targets, i.e.\ their varying distances
Across these configurations, we observe that the CNN maintains high performance overall target classification $75\%$ median accuracy, while decomposition into rows and columns yields $90\%$ accuracy for row prediction and $80\%$ for column prediction. 
These results demonstrate only a small improvement when decomposing the problem into separate row and column predictions. However, strategy 2 could be used to generalize the classification task into a precise endpoint prediction for both azimuth and altitude angle.
Last, there is a significant difference in prediction accuracy between row and column classification. This discrepancy is due to greater spatial dispersion of targets in the horizontal dimension compared to the vertical dimension due to the setting of the task.

%
%
%
%%%%%%%%%%%%%%%%%%%%%%%%%%%%%%%%%%%%%%%%%%%%%%%%
%\clearpage
\section{Discussion}\label{sec:Discussion}
Classification performance exhibits substantial inter-subject variability. While several participants achieve high accuracies, others show reduced performance. This high variability of EMG signal derives from differences in anatomy, electrode placement, and subcutaneous tissue properties \cite{sheng2019common}, \cite{farina2014extraction},  \cite{araujo2000inter}, \cite{nordander2003influence}, \cite{hogrel1998variability}, \cite{farina2002influence}, \cite{schmidt2023concepts}. Visual inspection of the corresponding EMG signals suggests that participants with lower predictive accuracy exhibited increased signal variability and noise, which may arise from differences in electrode placement, muscle activation strategies, or anatomical variability. Also, the trajectory of the arm movement varies more. Such inter-subject differences are well known in EMG-based classification studies and highlight the challenges of developing subject-independent or universally robust decoding models.

The channel selection analysis reveals physiologically plausible patterns. As expected, wrist muscles contribute less to classification performance, as their primary role in this task is limited to stabilizing and manipulating the handheld controller, while the reaching movement itself is dominated by proximal, gross motor actions of the upper limb. Additionally, the trapezius muscle shows lower importance, indicating it provides little information to differentiate between reach directions. The relevance of the biceps and triceps muscles is consistent with their role in elbow flexion and extension during arm transport. Similarly, the anterior deltoid and pectoralis major, both contributing to forward-directed shoulder motion, are among the most informative channels, which aligns with the biomechanics of reaching movements performed in front of the body. Overall, we identified $7$ out of $10$ channels having a strong contribution to differentiation between reach directions.

The feature selection results are consistent with previous EMG classification studies \cite{phinyomark2018feature},  \cite{corvini2025emg}. Time-domain features exhibited high discriminative power, reflecting their sensitivity to muscle activation amplitude and temporal structure. In addition, wavelet entropy emerged as a highly informative feature, capturing the complexity and non-stationary characteristics of EMG signals. The importance of correlation-based features is also mathematically sound, as coordinated muscle activation and co-contraction patterns inherently encode task-specific information.

Optimizing the number of channels, features, and temporal windows resulted in a substantial reduction of the input data dimensionality while maintaining classification performance. The best-performing configuration consisted of seven EMG channels, eight features, and a combination of seven temporal windows plus an entire window spanning the whole reach, achieving a median accuracy of $80\%$. Which is high compared to other reaching predictions \cite{irastorza2017design} for $8$ targets with around $70\%$. 
%Notably, using only the single most informative temporal window still yielded an accuracy of $75\%$, demonstrating that a compact representation can retain most of the discriminative information. This efficiency is particularly relevant for real-time applications with strict computational and latency constraints.

Furthermore, we investigate the earliest time point at which movement intention can be reliably decoded. As expected, prediction performance is highest when using the full reach. However, meaningful classification is already possible before initial target contact. During the pre-motion interval, when the target is known but movement execution has not yet started, classification accuracy reaches $13\%$, averaged for all $25$ targets, and $64\%$ in the simplified four-class scenario. This finding indicates that preparatory muscle activity encodes target-specific information well before actual movement begins. The ability to decode movement intention during pre-motion and early execution phases is particularly relevant for assistive robotics, exoskeleton control, and neurorehabilitation. Early intention detection could enable more responsive and anticipatory assistance, improving user comfort and task performance. Moreover, the demonstrated reduction in required sensors and features supports the feasibility of deploying such systems in practical, wearable settings.

Last, we validate our findings with a CNN on the EMG data instead of the RF on the pre-calculated features to achieve similar accuracy. The split between separate row and column prediction offers a broader applications for any task with underlying geometrical information and can also allow endpoint prediction.
%
%
%
%%%%%%%%%%%%%%%%%%%%%%%%%%%%%%%%%%%%%%%%%%%%%%%%
\section{Conclusion}
This study demonstrates that intended reaching targets can be reliably decoded from upper-limb EMG signals using a compact and physiologically meaningful representation. By systematically analyzing muscle contributions, feature relevance, and temporal segmentation, we identify a reduced configuration of seven EMG channels, eight features, and a limited number of temporal windows that achieves classification accuracies of up to $80\%$ with minimal loss in performance. The results show that proximal muscles involved in gross arm transport dominate target discrimination, while wrist muscles contribute little. Time-domain and wavelet-based features provide the most informative representations, consistent with the coordinated and non-stationary nature of EMG during reaching. Crucially, target intention can be predicted not only during movement execution but to a smaller amount also before movement onset, with performance remaining well above chance level in the pre-motion phase. This early predictability highlights the potential of EMG-based intention decoding for anticipatory control in assistive and rehabilitation technologies.

\subsubsection{Acknowledgment}
This work is supported by grants GE-2-2-023A (REXO) and IT-2-2-023 (VAFES), and grant 163V9070 (RooWalk-HRW).
\subsubsection{Disclosure of Interest}
The authors declare that they have neither a financial nor a non-financial competing interest.
\footnotesize
\bibliography{Literatur,zoteroGroup}

@article{schmidt2023concepts,
  title={The concepts of muscle activity generation driven by upper limb kinematics},
  author={Schmidt, Marie D and Glasmachers, Tobias and Iossifidis, Ioannis},
  journal={BioMedical Engineering OnLine},
  volume={22},
  number={1},
  pages={1--29},
  year={2023},
  publisher={BioMed Central}
}

@article{konrad2005emg,
  title={EMG-{F}ibel},
  author={Konrad, Peter},
  journal={Eine praxisorientierte Einf{\"u}hrung in die kinesiologische Elektromyographie},
  year={2005}
}

@article{sheng2019common,
  title={Common spatial-spectral analysis of EMG signals for multiday and multiuser myoelectric interface},
  author={Sheng, Xinjun and Lv, Bo and Guo, Weichao and Zhu, Xiangyang},
  journal={Biomedical Signal Processing and Control},
  volume={53},
  pages={101572},
  year={2019},
  publisher={Elsevier}
}

@article{farina2014extraction,
  title={The extraction of neural information from the surface EMG for the control of upper-limb prostheses: emerging avenues and challenges},
  author={Farina, Dario and Jiang, Ning and Rehbaum, Hubertus and Holobar, Ale{\v{s}} and Graimann, Bernhard and Dietl, Hans and Aszmann, Oskar C},
  journal={IEEE Transactions on Neural Systems and Rehabilitation Engineering},
  volume={22},
  number={4},
  pages={797--809},
  year={2014},
  publisher={IEEE}
}

@article{araujo2000inter,
  title={On the inter-and intra-subject variability of the electromyographic signal in isometric contractions},
  author={Araujo, R Correa and Duarte, M and Amadio, A Carlos},
  journal={Electromyography and Clinical Neurophysiology},
  volume={40},
  number={4},
  pages={225--230},
  year={2000},
  publisher={EDITIONS NAUWELAERTS SA}
}

@article{nordander2003influence,
  title={Influence of the subcutaneous fat layer, as measured by ultrasound, skinfold calipers and BMI, on the EMG amplitude},
  author={Nordander, Catarina and Willner, Julian and Hansson, G-{\AA} and Larsson, Britt and Unge, Jeannette and Granquist, L and Skerfving, Staffan},
  journal={European Journal of Applied Physiology},
  volume={89},
  number={6},
  pages={514--519},
  year={2003},
  publisher={Springer}
}

@article{hogrel1998variability,
  title={Variability of some SEMG parameter estimates with electrode location},
  author={Hogrel, J-Y and Duch{\^e}ne, Jacques and Marini, J-F},
  journal={Journal of Electromyography and Kinesiology},
  volume={8},
  number={5},
  pages={305--315},
  year={1998},
  publisher={Elsevier}
}

@article{farina2002influence,
  title={Influence of anatomical, physical, and detection-system parameters on surface EMG},
  author={Farina, Dario and Cescon, Corrado and Merletti, Roberto},
  journal={Biological Cybernetics},
  volume={86},
  number={6},
  pages={445--456},
  year={2002},
  publisher={Springer}
}

@article{corvini2025emg,
  title={EMG-Based reaching prediction for upper limb rehabilitation: a systematic analysis of factors affecting the classification accuracy},
  author={Corvini, Giovanni and de Nobile, Alessia and Del Grossi, Tommaso and De Marchis, Cristiano and Gandolla, Marta and Ambrosini, Emilia and Schmid, Maurizio},
  journal={Research Square},
  year={2025}
}

@manual{Warner2023ProCalc,
  title        = {Instructions for ProCalc Implementation of Upper Limb Model},
  author       = {Warner, Martin},
  year         = {2023},
  month        = jun,
  version      = {1},
  institution  = {School of Health Sciences, University of Southampton},
  address      = {Southampton, UK},
  note         = {Unpublished technical documentation},
}

@article{phinyomark2018feature,
  title={Feature extraction and selection for myoelectric control based on wearable EMG sensors},
  author={Phinyomark, Angkoon and N. Khushaba, Rami and Scheme, Erik},
  journal={Sensors},
  volume={18},
  number={5},
  pages={1615},
  year={2018},
  publisher={MDPI}
}

@article{Breiman2001RF,
  author  = {Breiman, Leo},
  title   = {Random Forests},
  journal = {Machine Learning},
  volume  = {45},
  number  = {1},
  pages   = {5--32},
  year    = {2001},
  doi     = {10.1023/A:1010933404324}
}

@inproceedings{irastorza2017design,
  title={Design of continuous EMG classification approaches towards the control of a robotic exoskeleton in reaching movements},
  author={Irastorza-Landa, Nerea and Sarasola-Sanz, Andrea and L{\'o}pez-Larraz, Eduardo and Bibi{\'a}n, Carlos and Shiman, Farid and Birbaumer, Niels and Ramos-Murguialday, Ander},
  booktitle={2017 International Conference on Rehabilitation Robotics (ICORR)},
  pages={128--133},
  year={2017},
  organization={IEEE}
}

@article{deecke1969distribution,
  title={Distribution of readiness potential, pre-motion positivity, and motor potential of the human cerebral cortex preceding voluntary finger movements},
  author={Deecke, L{\"u}der and Scheid, Peter and Kornhuber, Hans H},
  journal={Experimental brain research},
  volume={7},
  number={2},
  pages={158--168},
  year={1969},
  publisher={Springer}
}

@article{schmidt2026insights,
  title={Insights into motor control: predict muscle activity from upper limb kinematics with LSTM networks},
  author={Schmidt, Marie D and Glasmachers, Tobias and Iossifidis, Ioannis},
  journal={Scientific Reports},
  year={2026},
  publisher={Nature Publishing Group UK London}
}

@inproceedings{mora2021multi,
  title={Multi-subject identification of hand movements using machine learning},
  author={Mora-Rubio, Alejandro and Alzate-Grisales, Jesus Alejandro and Arias-Garz{\'o}n, Daniel and Buritic{\'a}, Jorge Iv{\'a}n Padilla and Var{\'o}n, Cristian Felipe Jim{\'e}nez and Bravo-Ortiz, Mario Alejandro and Arteaga-Arteaga, Harold Brayan and Hassaballah, Mahmoud and Orozco-Arias, Simon and Isaza, Gustavo and others},
  booktitle={Sustainable Smart Cities and Territories International Conference},
  pages={117--128},
  year={2021},
  organization={Springer}
}

@article{fernandez2014we,
  title={Do we need hundreds of classifiers to solve real world classification problems?},
  author={Fern{\'a}ndez-Delgado, Manuel and Cernadas, Eva and Barro, Sen{\'e}n and Amorim, Dinani},
  journal={The journal of machine learning research},
  volume={15},
  number={1},
  pages={3133--3181},
  year={2014},
  publisher={JMLR. org}
}

\end{document}